\documentstyle[aps,prl,twocolumn]{revtex}
\begin{document}
\draft
\wideabs{
\title{Spin Fluctuation and Persistent Current in a Mesoscopic Ring 
       Coupled to a Quantum Dot}
\author{Sam Young Cho$^{1,2}$, Kicheon Kang$^3$, 
        Chul Koo Kim$^{1,2}$, and Chang-Mo Ryu$^4$}
\address{ $^1$Institute of Physics and
          Applied physics, Yonsei University, Seoul 120-749, Korea}        
\address{ $^2$Center for Strongly Correlated Materials Research,
          Seoul National University, Seoul 151-742, Korea}
\address{ $^3$Department of Physics, Chonbuk National University,
           Chonju 561-756, Korea}
\address{ $^4$Department of Physics, 
          Pohang University of Science and Technology,
          Pohang 790-784, Korea}
\date{\today}

\maketitle
\begin{abstract}
 We investigate the persistent current influenced by the spin fluctuations
 in a mesoscopic ring weakly coupled to a quantum dot.
 It is shown that the Kondo effect gives
 rise to some unusual features of the persistent current
 in the limit where the charge transfer between two subsystems is suppressed.
 Various aspects of the crossover 
 from a delocalized to a localized dot limit 
 are discussed in relation with the effect of
 the coherent response of the Kondo cloud to the Aharonov-Bohm flux.
\end{abstract}

}

  Coherent charge transfer from one region of a system into another 
  in a coupled mesoscopic structure can dramatically change 
  the physical properties of the coupled system. 
  A typical example is a mesoscopic ring coupled to 
  a finite-size wire\cite{stub_pc} or weakly connected to
  a quantum dot(QD)\cite{Buttiker96,Cedraschi}.
  If the charge transfer between the two subsystems is suppressed,
  these two regions are effectively decoupled from each other 
  in ordinary cases.  At very low temperature, however, 
  the spin degree of freedom may play a crucial role
  in determining the ground state(GS) properties of the coupled system 
  even when the charge fluctuations are suppressed. 
  A system composed of a metallic host  and a magnetic impurity 
  offers a good opportunity to study such type of effects.
  Actually, 
  the Kondo problem, which was first investigated
  in dilute magnetic alloys\cite{Hewson},
  provides a best known example.
  Recent experimental efforts have enabled the realization of the Kondo effect
  in the semiconductor-based QDs~\cite{Gordon98,Cronen98,Schmid98,Simmel99}.
  In the unitary scattering limit, the Kondo effect gives rise to a perfect
  transmission\cite{Glazman88,Ng88,Wiel} through the dot.
  The perfect transmission can be interpreted in terms of  
  a Kondo cloud originating from 
  the itinerant electrons in the leads.
  Furthermore, using the Aharonov-Bohm(AB) interferometers,
  phase-coherent Kondo effect has been reported
  and discussed in Refs.\cite{Wiel,Gerland,Ji}.
  However, transport measurements for such open systems
  do not provide clear understanding on the phase-coherent property
  of the GS in closed systems.
  On the other hand, 
  persistent current(PC) 
  in an isolated AB ring with
  a coupled dot provides useful information about the effect of
  spin fluctuations on the GS,
  which cannot be addressed in open systems.
  There have been some previous works on the Kondo-assisted PC  
  in a QD embedded in an AB ring\cite{Ferrari99,Kang99}.
  
\begin{figure}
\vspace*{5.0cm}
\includegraphics{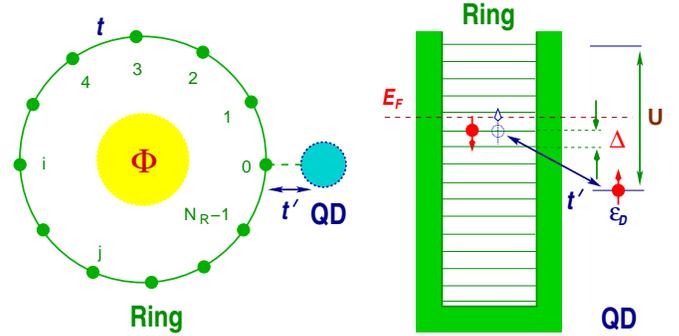}
\caption{Left : An Aharonov-Bohm ring threaded by the magnetic flux $\Phi$
and laterally coupled to a quantum dot.
The coupling between the dot and the site ``0" of the ring
is characterized by a tunnel matrix element $t'$.
Right : Schematic energy diagram of the system.}
\label{fig:model}
\end{figure}

  In this Letter, we investigate the equilibrium properties of
  a QD side-attached to a mesoscopic ring threaded by
  an AB flux(see Fig.\ref{fig:model}).
  Here, 
  we focus on the effect of spin fluctuations that strongly
  affect the PC and their phase-coherent properties 
  which are functions
  of the level spacing, $\Delta$, and the total number of electrons, $N_e$. 
  We show that the crossover from the strongly coupling 
  to a localized dot limit introduces various interesting phenomena. 

  The Hamiltonian of our coupled system 
  is decomposed into the three parts as 
  \begin{equation}
  H = H_R + H_D + H_T,
  \end{equation}
  where $H_R$, $H_D$, and $H_T$ represent the AB ring, the QD, and
  the tunneling interaction between the ring and the dot, respectively.
  The ring is described by an ideal one-dimensional tight-binding model
  with $N_R$ lattice sites; 
  \begin{equation}
    H_R = -t \sum^{N_R-1}_{j=0}\sum_{\sigma}
     \left( e^{i2\pi\phi/N_R} c^\dagger_{j+1,\sigma} c_{j,\sigma}
            + {\rm H.C.} \right),
  \end{equation}
  where $t$ denotes the hopping integral between the neighboring sites.
  $H_R$ can be diagonalized with the eigenvalues
  $\varepsilon_m = -2 t \cos[(2\pi/N_R)(m+\phi)]$, 
  where $m$ is an integer 
  within the interval $[-(N_R-1)/2,N_R/2]$
  and  $\phi=\Phi/\Phi_0$ with $\Phi$ and $\Phi_0(=hc/e)$ being
  the magnetic flux and the flux quantum, respectively.
  The QD, which has a much smaller size compared to the ring,
  can be described by a spin-degenerate single level 
  with a strong on-site Coulomb repulsion $U$.
  At the site ``0''of the ring,
  electrons are allowed to hop to the dot or vice versa.
  These processes are characterized by a tunnel matrix element $t'$.
  In the present study, 
  we adopt the slave-boson(SB) representation for infinite $U$ 
  in describing the low energy physics in the spin fluctuation limit.
  In this representation, 
  the electron annihilation operator in the dot, $d_\sigma$,
  consists of the SB operator $b^\dagger$ which creates
  an empty state and a pseudo-fermion operator $f_\sigma$ which
  annihilates the singly occupied state with energy $\varepsilon_D$ and
  spin $\sigma$ in the dot; $d_\sigma = b^\dagger \, f_\sigma$.
  The Hamiltonians for the dot and for the coupling are given 
  in this representation by 
   \begin{mathletters}
   \begin{eqnarray}
   H_D &=& \sum_{\sigma} \varepsilon_D f^\dagger_\sigma f_\sigma, \\
   H_T &=& - t' \sum_{\sigma} 
      \left( c^\dagger_{0,\sigma} \, b^\dagger \, f_\sigma
         + f^\dagger_\sigma \, b \, c_{0,\sigma} \right) ,
   \end{eqnarray}
   \end{mathletters}
   respectively.
  The infinite $U$ limit gives rise to the constraint
  preventing double occupancy in the dot. This emerges in
  the pseudo-charge constraint $Q=1$, where
  $Q = b^\dagger b + \sum_\sigma f^\dagger_\sigma f_\sigma$.
  One way to impose the constraint is to add a ``chemical potential"
  term $\lambda Q$ to $H$, and to project it onto the physical
  subspace $Q=1$ by taking $\lambda \rightarrow \infty$ at the end of
  calculation\cite{Coleman}.
 
  The total Hamiltonian is transformed to the following form
  by using the diagonalized basis of $H_R$ as 
  \begin{eqnarray}
  H&=&\sum_{m\sigma} \varepsilon_m c^\dagger_{m\sigma} c_{m\sigma}
    + \sum_\sigma \varepsilon_D f^\dagger_\sigma f_\sigma
  \nonumber \\ &&
    + \sum_{m\sigma} \left( t_m c^\dagger_{m\sigma}\, b^\dagger\, f_\sigma
        + t^*_m f^\dagger_\sigma \, b \, c_{m\sigma} \right) ,
  \end{eqnarray}
  where $t_m =-t'/\sqrt{N_R}$.
  The effective width of the dot level due to the coupling
  between the dot and the ring is given by
  $\Gamma = \pi {\cal N}_D(\varepsilon_F) t^2_m(\varepsilon_F)$.
  Here, ${\cal N}_D(\varepsilon_F)=1/\Delta$ is the density of states (DOS) 
  at the highest occupied level energy, $\varepsilon_F$, of the ring. 
  For half-filling $N_R=N_e$, the effective width becomes
  $\Gamma= \pi |t'|^2/(2tN_e\sin\left[\pi/N_e\right])$.

  The partition function $Z$ of the $N_e$-electron system
  of the Hamiltonian is shown to be expressed as a product $Z=Z_{0} Z_{QD}$,
  where $Z_0$ is the partition function of the $N_e$ free 
  electrons in the ring
  and $Z_{QD}$ corresponds to the contribution from the interacting QD; 
  \begin{equation}
  Z_{QD} = \int^\infty_{-\infty} d\varepsilon \; e^{-\beta \varepsilon}
  \Big( \varrho_b(\varepsilon) + \sum_\sigma \varrho_f(\varepsilon) \Big).
  \end{equation}
  The spectral functions $\varrho_\alpha(\varepsilon)(\alpha=b,f)$ 
  for pseudo-particles are given by imaginary part of
  the corresponding retarded Green's functions;
  $\varrho_\alpha (\varepsilon)
      =  -(1/\pi) {\rm Im} \, {\cal G}_\alpha (\varepsilon)$.

  We adopt the leading order $1/N_s$-diagrammatic expansion
  with $N_s$ being the magnetic degeneracy.
  It is well known that this approximation describes well the essence 
  of the Kondo correlation preserving the Fermi liquid properties\cite{Hewson}. 
  In the leading order, the Green's functions of the pseudo-particles
  are given by 
  \begin{mathletters}
  \begin{eqnarray}
  {\cal G}_f(\varepsilon)
    &=& \frac{1}{ \varepsilon - \varepsilon_D +i0^+}, \\
  {\cal G}_b(\varepsilon)
    &=& \frac{1}{ \varepsilon - \Pi_0(\varepsilon) +i0^+}.
  \end{eqnarray}
  \end{mathletters}
  The self-energy of the SB\cite{Kang96,Matsuura} is given by 
  \begin{equation}
  \Pi_0(\varepsilon) = \sum_{m\sigma} f(\varepsilon_{m})
   \frac{ \mid t_m \mid^2} {\varepsilon - \varepsilon_D + \varepsilon_m},
  \end{equation}
  where $f(\varepsilon_{m})$ is the occupation probability of the level $m$.
  This leads to the spectral function of the SB of the form 
  \begin{equation}
  \varrho_b (\varepsilon)
     = {\cal Z}(E_0) \delta(\varepsilon-E_0),
  \end{equation}
  where the {\it renormalization factor} ${\cal Z}(E_0)$ is given by
  \begin{equation}
   {\cal Z}(E_0)
     = \left( 1 -\frac{\partial}{\partial \varepsilon} \Pi_0(\varepsilon)
       \right)^{-1} \Bigg|_{\varepsilon=E_0}.
  \end{equation}
  $E_0$ is  obtained by  the self-consistent equation;
  $E_0 = \Pi_0(E_0)$, and corresponds to the GS energy
  for $T=0$.

  The intrinsic characteristic energy scale of the system, $T^0_K$,
  can be defined by the difference of the GS energy
  and the lowest excited state energy, in the bulk limit of the ring :
  \begin{equation}
   T^0_K \equiv \varepsilon_D -\varepsilon_F -E_0. 
  \end{equation}
  This corresponds to the bulk Kondo temperature in our scheme.

  The persistent current is defined by the relation
  \begin{equation}
   I(\phi,T) =  -\frac{e}{h} \frac{\partial}{\partial \phi}  F(\phi,T),
  \end{equation}
  where the free energy is given by $F = -(1/\beta) \ln Z$.
  The current can be decomposed into two parts; 
  $I(\phi,T)=I_0(\phi,T)+I_{QD}(\phi,T)$.
  $I_0(\phi,T)$ denotes the current contribution of the $N_e$ 
  free electrons in the ring which can be expressed as
  \begin{equation}
   I_0(\phi,T) =\sum_{m\sigma} f(\varepsilon_{m}) I_m(\phi) .
  \end{equation}
  Contribution from each level is given by 
  $I_m(\phi)=-(2e/h)\Delta \sin[(2\pi/N_R)(m+\phi)]$
  with the oscillation amplitude, $\Delta=2\pi t/N_R$.
  Note that $\Delta$ corresponds to the mean level spacing 
  in the bulk limit.
  It has been shown by Loss and Goldbart\cite{Loss91} that  
  the behavior of $I_0$ for an ideal ring 
  depends on the number of electrons with modulo 4.
  That is, there exist four different 
  number classes {\em i} $(i=0,1,2,3)$
  for $I^{N_e}_0$ with $N_e$ being defined as
  $N_e=4n-i$($n$; any positive number).

  At low temperature, $T \ll T^0_K$,
  the PC circulating in the ring of the system
  has the following form;
  \begin{equation}
   I(\phi,T) = \sum_{m\sigma}  {\cal F}(\varepsilon_m) I_m(\phi).
  \end{equation}
  The effective occupation probability in the state $m$,
  ${\cal F}(\varepsilon_m)$, is given by
  \begin{equation}
   {\cal F}(\varepsilon_m) = f(\varepsilon_m)
   \left(1 -{\cal Z}(E_0) \frac{\mid t_m \mid^2}
           { \left( \varepsilon_m + E_0 - \varepsilon_D \right)^2} \right).
  \label{eq:distribution}\end{equation}
  Several points can be clarified in terms of Eq. (\ref{eq:distribution}).
  In the empty dot limit ($\varepsilon_D-\varepsilon_F \gg 2\Gamma$ and 
  $n_D\rightarrow0$ with $n_D$ being the occupation number of the dot),
  the second term of Eq. (\ref{eq:distribution}) is negligible, and
  $I$ is equivalent to $I_0$.
  In other words, the two mesoscopic regions are effectively decoupled,
  because charge transfer is absent.
  However, when $\varepsilon_D$ is lowered to a charge fluctuation regime
  ($|\varepsilon_D-\varepsilon_F| \lesssim 2\Gamma$), the second term
  is no longer negligible and gives rise to contributions to the PC 
  induced by the charge transfer. 
  Such an effect of the charge transfer in this geometry
  were discussed by B\"uttiker and Stafford\cite{Buttiker96}. 

\begin{figure}
\vspace*{7.0cm}
\includegraphics{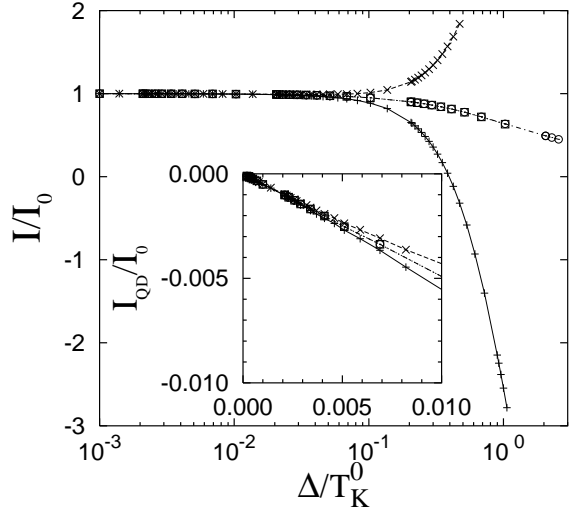}
\caption{Normalized persistent currents as a function of the energy level 
 spacing of the ring for $\phi=0.23$ and $T=0$.  
 The other parameters are chosen as $\varepsilon_D = -0.75$, $t'=\sqrt{0.3}$,
 and $t=1.0$ ($\Gamma \simeq 0.15$). 
 With these parameters, the Kondo correlation energy is achieved 
 at $T^0_K = 0.16 \times 10^{-3}$.
 The currents for different number classes are denoted by 
 $0(\odot)$,$1(+)$,$2(\Box)$, and $3(\times)$, respectively.
 In the inset, 
 the dot contribution of the current is shown in a linear scale. }
\label{fig:scale}
\end{figure}

  Here, we are mainly interested in the Kondo limit
  ($\varepsilon_D \! < \! -2 \Gamma$, and $n_D \rightarrow 1$)\cite{Mahan} 
  where the charge transfer is suppressed 
  and the spin fluctuation plays a crucial role.
  Without loss of generality,
  we choose the parameters $\varepsilon_D = -0.75$, and $2\Gamma \simeq 0.30$.
  In fact, 
  the parameters, as long as they are chosen to suppress 
  the charge fluctuation, does not
  affect the physics in the Kondo limit
  but only shifts 
  the value of the characteristic energy scale $T^0_K$.
  For simplicity, we consider only the half-filled case 
  : $N_e=N_R$. 
  Figure \ref{fig:scale} shows the numerical result of the PC 
  as a function of the normalized level spacing, $\Delta/T_K^0$.
  The behavior of $I/I_0$ depends strongly 
  on which number class $N_e$ belongs to,
  aside from the class dependence of $I_0$ itself which has been
  discussed in Ref. \cite{Loss91}.
  The Kondo effect manifested by
  the second term of Eq. (\ref{eq:distribution}) 
  is expected to
  provide a significant modification  on the current. 
  It is instructive to consider the following simple argument:
  Since the occupation of the dot goes to unity in the Kondo limit, 
  the dot has one electron and the ring $N_e-1$ electrons.
  If these two systems are effectively decoupled,
  the PC of the coupled $N_e$ electron system
  should be equivalent to that
  of an ideal ring with $N_e-1$ electrons.
  Thus, one may expect that $I^{N_e} \simeq I^{N_e-1}_0$.
  However, our result demonstrates that
  $I^{N_e}=I^{N_e}_0$ in the limit of $\Delta/T^0_K \rightarrow 0$,
  which is in agreement
  with the recent result obtained by an exact Bethe ansatz
  calculation\cite{Eckle}.
  It is interesting to note that the Kondo impurity does not affect 
  the PC in the continuum limit
  in spite of the fact that
  the dot captures one electron from the ring.
  This result is interpreted as follows. 
  About $N_{eff}\approx T_K^0/\Delta$ electrons in the ring take part
  in forming a Kondo screening cloud\cite{Bergmann}. 
  Our result in the continuum limit shows
  that the screening cloud exactly compensates 
  the single trapped electron in the dot for a response to the AB flux.
  In other words, the Kondo cloud plays a role of an extra electron
  in the ring which participates in the coherent motion.

  Our conclusion for the continuum limit
  may appear in conflict with 
  the result of the electron transmission
  in a quantum wire with a side-coupled quantum dot\cite{Kang00}. 
  The electron transmission in this open system is shown to be completely
  suppressed due to a destructive interference between 
  the ballistic 
  and the Kondo channel. 
  We believe that this discrepancy comes from the difference
  between a closed and an open system.
  In contrast to the current response to the
  applied voltage in a quantum wire,
  the response direction of the PC to the magnetic flux
  in a ring depends on the symmetry of each eigenstate of the ring.
  As a result of the alternating sign of the response, the QD makes no 
  net effect on the PC in the limit of $\Delta/T_K^0\rightarrow0$. 

  As shown in Fig. \ref{fig:scale}, the effect of finite 
  level spacing is to reduce the PC by the amount
  of ${\cal O}(\Delta/T^0_K)$ in the $\Delta/T_K^0 \ll1$ limit. 
  The inset of Fig. \ref{fig:scale} shows 
  that $I_{QD}/I_0 = -\alpha(\Delta/T^0_K)$ at very small $\Delta/T^0_K$.
  Note that the numerical coefficient 
  $\alpha\approx0.5$ is independent of the number class and $\phi$.
  While this reduction of the current becomes non-linear
  for odd $N_e$ as $\Delta/T_K^0$ increases,
  the linear scaling is preserved for a wider range of $\Delta/T^0_K$
  for $N_e=\mbox{\rm even}$. Actually, 
  the reduction of the PC 
  for finite $\Delta$ indicates a weakening 
  of the Kondo screening due to the finite size of the metallic host.

  To understand the linear scaling behavior of $I_{QD}/I_0$ 
  more clearly, 
  it is helpful
  to define the effective ``charge" for each level of the ring, 
  $Q_K^{m\sigma}$, which take parts in forming the Kondo cloud:
  \begin{equation}
   Q^{m\sigma}_K/e \equiv
   -{\cal Z}(E_0) \frac{|t_m|^2}{(\varepsilon_m+E_0-\varepsilon_D)^2}.
  \label{eq:charge}\end{equation}
   It should be noted that the net contribution 
   of the effective charge,  
  \begin{equation}
  \sum_{m\sigma} Q^{m\sigma}_K/e = -n_D, 
  \end{equation}
  corresponds to the {\em hole} that arises 
  from the local charge trapped by the QD.  
  In the Kondo limit of $n_D=1$, 
  one can find that
  \begin{equation}
   Q_K^{m\sigma}/e \sim -\Delta/T_K^0  
  \end{equation}
  for $ -T_K^0 \lesssim \varepsilon_m-\varepsilon_F$, and
  $Q_K^{m\sigma}/e \simeq 0$ for $\varepsilon_m-\varepsilon_F \ll -T_K^0$.
  Therefore, $I_{QD}$ can be 
  estimated as
  \begin{mathletters}
  \begin{eqnarray}
   I_{QD}&=&\sum_{m\sigma} 
            f(\varepsilon_m) \left(\frac{Q_K^{m\sigma}}{e}\right) I_m \\
   &\sim& -\frac{\Delta}{T_K^0} \sum_{m\sigma} f(\varepsilon_m) I_m 
     = -\frac{\Delta}{T_K^0} I_0 ,
  \end{eqnarray} 
  \end{mathletters}
  which explains the linear scaling of $I_{QD}/I_0$.

  Further increase of $\Delta$ results in a various kinds of 
  crossover behaviors in $I/I_0$ around 
  $0.05 \lesssim \Delta/T^0_K \lesssim 1.0$,
  as shown in Fig. \ref{fig:scale}.
  For a given AB flux ($\phi=0.23$), $I^{N_e}/I^{N_e}_0$ for the class 3
  rapidly increases, whereas its direction is reversed for the class 1.
  In contrast, 
  the current displays a slow crossover for even $N_e$. 
  Note that the behavior of the crossover depends not only 
  on the number class but also on the value of $\phi$.
  However, in general, the crossover can be
  explained in terms of a continuous evolution from 
  a delocalized Kondo
  to a localized dot limit. 
  The crossover can be clearly seen by considering 
  the limit of very large level spacing, $\Delta/T^0_K \gg 1$.
  In this limit, only the topmost level of the ring
  contributes to the modification of the current, and one can write as
  $I^{N_e}_{QD}(\phi)=-n_D I_F(\phi)$, where $I_F$ denotes the current
  component of the highest occupied level, $m=F$.
  Since $n_D \simeq 1$ in our case, we obtain the relation
  \begin{equation}
   I^{N_e}(\phi) \simeq I^{N_e-1}_0(\phi).
  \end{equation}
  This result is exactly what one would expect 
  for an effectively decoupled system. 
  Therefore, as $\Delta/T^0_K$ increases, the behavior of the current
  evolves from $I^{N_e}=I^{N_e}_0$ of the delocalized dot limit
  to $I^{N_e}=I^{N_e-1}_0$ of the localized dot limit.

  The number class dependence of the PC for an ideal ring\cite{Loss91}
  allows to understand the PC behavior of
  our system as a function of $\Delta/T^0_K$
  for a given AB flux.
  For instance, in the range of $0<\phi<0.25$,
  class 1 undergoes a transition to class 2 as $\Delta/T^0_K$
  is increased. 
  This changes the direction of the current,
  since $I^{N_e-1}_0/I^{N_e}_0 < 0$ in this region of $\phi$.
  Class 3 
  transforms into class 0 with an enhanced value of 
  $I/I_0$ since $I^{N_e-1}_0/I^{N_e}_0 > 1$. All the other cases
  for any number class and $\phi$ can be understood in the same way. 
  

  We have investigated the effect of spin fluctuations 
  on the equilibrium properties of a coupled mesoscopic ring-quantum 
  dot system.
  The phase-coherent response of the strongly correlated ground state
  to the AB flux was shown to depend on the ratio
  of the finite level spacing of the ring to the Kondo temperature.
  A crossover 
  from a strongly coupled to effectively decoupled system
  provides several unusual 
  features in the behavior of the persistent current.

 This work was partially supported by
 the Korea Research Foundation(99-005-D00011).
 K. Kang was supported by the Basic Research Program of the KOSEF
 (1999-2-11400-005-5). 


\end{document}